\documentclass[twocolumn,showpacs,preprintnumbers,amssymb]{revtex4}
\usepackage{amsmath,amstext,amsthm,enumerate}
\usepackage[dvips]{graphicx}
\usepackage{amsfonts}
\usepackage{amsmath}
\def\a{{\alpha}}

\def\lam{{\lambda}}

\def\h{\hbar}

\def\RR{{\mathbb R}}

\def\NN{{\mathbb N}}

\newcommand{\ket}[1]{\ensuremath{\left|#1\right\rangle}}

\newcommand{\erf}[1]{\ensuremath{\mathrm{erf}\:#1}}
\newcommand\beq{\begin{equation}}
\newcommand\eeq{\end{equation}}
\newcommand\bsp{\begin{split}}
\newcommand\espl{\end{split}}
\newcommand{\dep}[2]{\ensuremath{\frac{\partial #1}{\partial #2}}}

\theoremstyle{plain}

\newtheorem{thm}{Theorem}

\newtheorem{lem}{Lemma}

\theoremstyle{definition}

\newtheorem{defin}{Definition}

\newtheorem{rem}{Remark}

\begin{document}

\title{Superintegrability, isochronicity, and quantum harmonic behavior}

 \author{Simon Gravel}

 \affiliation{ CRM, Universit\'e de Montr\'eal, C.P.\-6128, Succursale
 Centre-Ville, Montr\'eal, Qu\'ebec,  H3C 3J7,
 Canada}

\begin{abstract}
We discuss the properties of superintegrable Hamiltonian systems,
in particular those that admit separation of variables in
cartesian coordinates. We show that the superintegrability of such
potentials is equivalent to the isochronicity of the separated
potentials. We use this fact to get a new insight into an old
question about the relation between quantum and classical harmonic
behavior.
\end{abstract}
 \maketitle

Hamiltonians of the form \beq \label{hsep}
H=p_x^2/2+p_y^2/2+V_a(x)+V_b(y)\eeq admit two independent
second-order integrals, $H_a=p_x^2/2+V_a(x)$ and
$H_b=p_y^2/2+V_b(y).$ They are therefore integrable. Those that
admit an additional nontrivial integral of motion are maximally
superintegrable since they have $2n-1$ integrals of motion, $n$
being the number of degrees of freedom. Such systems were
investigated in \cite{GW, Gr,WSUF}.


Theorem \ref{thma} relates the existence of this additional
integral of motion to the isochronicity of the separated
Hamiltonians. It provides a new characterization of isochronous
potentials in classical mechanics, which can easily be adapted to
many more general contexts.
We use this to propose a new definition of quantum harmonic
behavior, based on quantum superintegrability. We then give
examples that show that potentials that are harmonic according to
this definition do not necessarily have equidistant spectra, which
was the usual definition for quantum harmonic behavior, but still
exhibit regular behavior. The spectra of our examples can indeed
be generated from a finite number of states by the application of
a creation operator. Therefore their emission spectrum is highly
degenerate, which makes them physically very similar to the usual
harmonic oscillator.

\begin{rem}
We will assume through this paper that all classical potentials
are $C_2$ over the considered interval of the real axis. The
results might be generalizable to $C_1.$
\end{rem}

\begin{defin}
A Hamiltonian $H_a(x,p_x)$ and, by extension, the potential
$V_a(x)$ is isochronous if $V_a$ has a local minimum and if all
its bounded trajectories have the same period.
\end{defin}

\begin{rem}
Since the potentials considered here are differentiable, and the
existence of a local maximum would imply trajectories with
infinite period, an isochronous potential can have only one local
minimum, and the potential must go to infinity on the boundaries
of the considered domain.
\end{rem}






\begin{thm}\label{thma}
A Hamiltonian of the form \eqref{hsep} with a local minimum is
maximally superintegrable if and only if $V_a$ and $V_b$ are
isochronous.
\end{thm}

\begin{proof}
The first part of the proof is straightforward. We know, from a
theorem due to Nekhoroshev \cite{Ne}, that the bounded
trajectories of a maximally superintegrable Hamiltonian are all
closed. Since the motion along the $x$ and $y$ axis can be
decoupled, we can consider separately the periods along each
direction. If these periods varied, they would do it continuously
with respect to initial conditions that are independent (i.e. the
$x_i(0)$ and $p_i(0)$). We could therefore choose initial
conditions such that the ratio between the periods is irrational.
In this case the two-dimensional motion would not be periodic,
leading to a contradiction.

We now have to demonstrate that the converse is true, i.e. that if
$V_a$ and $V_b$ are isochronous the whole system is
superintegrable. The idea of the proof is to transform each
one-dimensional system into a harmonic oscillator, and then use
the two-dimensional integrals of the anisotropic harmonic
oscillator to find, via the inverse transformation, integrals of
the initial system.
We will find directly the appropriate transformation using the
trajectories of the system.

We can specify a point in the phase space of a one dimensional
isochronous potential that has its minimum at $x=0$ by specifying
the trajectory on which it lies, and the time it took for a
particle moving along that trajectory to get there starting from
the turning point with $x>0.$ Since there is only one such turning
point, and since the system is isochronous, this transformation is
well defined and one-to-one on the interval $t\in
\left[0,T\right),$ $T$ being the period of the system.

We therefore consider the transformation $(x,p)\rightarrow (r,t),$
where $r$ is the value of the turning point with $x>0$ and $t$ the
time for a particle to get to $(x,p)$ from that turning point. We
will now introduce a lemma that will be demonstrated below.

\begin{lem}\label{diffeo}
The transformation $(x,p)\rightarrow (r,t)$ is a diffeomorphism
for $r\neq 0$.
\end{lem}

We can compose this change of variables with

\beq X=r \cos t, P=r\sin t,\eeq

\noindent which is a transformation from polar to cartesian
coordinates and is continuous and differentiable, except at the
origin. The transformation $(x,p)\rightarrow (X,P)$ is therefore a
diffeomorphism, except possibly at the origin.

The trajectories in the $(X,P)$ plane have constant velocity,
period $T$ and are along circles. Therefore $X$ and $P$ obey the
equations of a harmonic oscillator,

\beq\bsp \dot X =P \hspace{1.5 cm} \dot P= -\left(\frac{2
\pi}{T}\right)^2 X
\end{split}\eeq

We use this method for both $(x_1,p_1)$ and $(x_2,p_2),$ and since
the periods in the two directions are commensurable, the resulting
system expressed in terms of $\{X_1,X_2,P_1,P_2\}$ exhibits the
same motion, and therefore the same integrals as the anisotropic
harmonic oscillator with rational ratio. This is known \cite{JH}
to admit three independent integrals of motion, $\{Q_1,Q_2,Q_3\}.$
%
%

Since the $X_i$ and $P_i$ are differentiable functions of the
$x_i$ and $p_i,$ The three $Q_i,$ expressed in terms of the
$\{x_i,p_i\}$ are independent integrals of motion of the initial
Hamiltonian.

\end{proof}


Let us now prove lemma \ref{diffeo}.
\begin{proof} 
Differentiability of $(x,p)$ with respect to time is
given by the Hamilton equations, $\dot x=p$ and $\dot p=-V'(x),$
%
%



%
%




The existence of the two derivatives with respect to $r$ is given
by the theorem of differentiable dependance on initial conditions
(see e.g. \cite{Ha}), applied to the Hamilton equations.

In order to show that $(x,p)\rightarrow (r,t)$ is also
differentiable, we have to show that the determinant of the
Jacobian matrix \begin{equation*} J=\left|\begin{array}{cc}
\dep{x}{t}&\dep{x}{r}\\\dep{p}{t}&\dep{p}{r}
\end{array}\right|=\left|\begin{array}{cc}
p(t,r)&\dep{x}{r}\\-V'(x(t,r))&\dep{p}{r}
\end{array}\right|\end{equation*}

\noindent does not vanish on a solution
$\left(x(t,r),p(t,r)\right)$. Let us write
\begin{equation*}
x(t,r+v)=x_v,\;\;\;p(t,r+v)=p_v,\;\;\;x_0=x,\;\;\;p_0=p.
\end{equation*}

The Jacobian reads
\begin{equation*}\bsp p\dep{p}{r}+V'(x)\dep{x}{r}=\lim_{v\rightarrow0}
p(p_v-p)/v+V'(x)(x_v-x)/v\\
=\lim_{v\rightarrow0}\left(p_v^2/2-p^2/2+V'(x)\left(x_v-x\right)\right)/v\\
-\lim_{v\rightarrow0}(p_v-p)^2/(2v)
\end{split}
\end{equation*}

Since $V'(x)(x_v-x)\simeq V(x_v)-V(x)$ and $(p_v-p)^2/v\simeq
(\partial{p}/\partial{r})^2 v \simeq 0$ we have
\begin{equation*}\bsp J= \lim_{v\rightarrow0}(p_v^2/2+V(x_v)-(p^2/2+V(x)))/v\\
=\lim_{v\rightarrow0}(V(r)-V(r+v))/v=-V'(r).\end{split}
\end{equation*}
This is nonzero everywhere except at $r=0$. The proof of the lemma
is complete.
\end{proof}

We now introduce a concept which will allow us to discuss
one-dimensional potentials without always referring to the
two-dimensional Hamiltonian.
\begin{defin}
A Hamiltonian $H_a=p_x^2/2+V_a(x)$ is called $2D$-superintegrable
if there exists a $H_b=p_y^2/2+V_b(y)$ such that $H_a+H_b$ is
superintegrable. We shall call $H_b,$ Hamiltonian associated to
$H_a.$ It is also $2D$-superintegrable.
\end{defin}

\begin{rem}
If $H_a$ is isochronous, theorem \ref{thma} guarantees that it is
$2D$-superintegrable. The fact that $H_a$ is $2D$-superintegrable
gives no guarantee that it is isochronous, though, unless both
$H_a$ and its associated Hamiltonian have a local minimum.
\end{rem}

$2D$-superintegrability is therefore a generalization of
isochronicity since it includes potentials such as free motion and
the repulsive harmonic oscillator. It is therefore tempting to
identify as "harmonic" $2D$-superintegrable potentials in
classical mechanics.

Even though some questions regarding the independence of quantum
integrals are not yet fully understood \cite{We, Ht3}, the concept
of superintegrability can be adapted to quantum mechanics in a
straightforward manner, and this had been done since \cite{FMSUW}.
The adaptation is not so easy for isochronicity. Since well-known
examples (such as the harmonic oscillator) have equidistant
spectra, it was natural to assume that this property should
characterize quantum harmonic potentials \cite{GH,Mo,NG}. It was
shown in \cite{GH,Mo} that the quantum equivalent to a classical
isochronous potential need not be equidistant. We will see that
this is still the case when we ask that the quantum equivalent be
$2D$-superintegrable. The quantum potential will nevertheless
exhibit properties that make it similar to the harmonic
oscillator, e.g. regarding the infinite degeneracy of the emission
spectrum. Therefore it is tempting to identify quantum harmonic
behavior and $2D$-superintegrability. Studying the properties of
$2D$-superintegrable potentials will then help us in the search
for a more "property-oriented" definition for quantum harmonic
behavior.

The formalism of creation-annihilation operators, defined here as
operators $A$ such that $[H,A]=\lam A$ with $\lam$ a constant, is
useful in dealing with $2D$-superintegrable potentials. Let us
write $H=H_a(x_1,p_1)+H_b(x_2,p_2),$ and let $Q(x_1,x_2,p_1,p_2)$
be the additional integral of motion. If the commutator $[H_a,Q]$
is nonzero, it is a new integral of $H.$ This is likely to be a
consequence of the definition of the independence of quantum
integrals since it is the case in classical mechanics. Therefore
the operator $M:Q\rightarrow [H_a,Q]$ is a nontrivial and nonzero
linear operator on the vector space of the integrals of motion of
$H.$ Therefore each nonzero eigenvalue of $M$ indicates the
existence of a creation or annihilation operator for $H_a.$ Notice
that the existence of creation-annihilation operators does not
guarantee that the spectrum is equidistant. Consider for example
the Hamiltonian $H=p^2/2+ax^2+b/x^2,$ which is classical and
quantum $2D$-superintegrable for all values of $a$ and $b.$ Its
spectrum and eigenfunctions are known \cite{WSUF}. In the quantum
case we can find a creation-annihilation pair by the method just
described, with $|\lam|=2 \sqrt{2a} \h.$

For $b>3/8 \h^2,$ the spectrum of this potential is equidistant.
For $- \h^2/8<b<3\h^2/8,$ though, the spectrum is given by
$$E= \sqrt{2 a} (2 k+1\pm \nu)\h,$$
\noindent where $\nu=\sqrt{1+8b/\h^2}/2.$ Therefore, apart from
the special case $b=0,$ the potentials with $-1/8\h^2<b<3/8\h^2,$
have a spectrum organized in pairs separated by $2\sqrt{2a
}\nu\h.$ Each pair is separated by $2 \sqrt{2a} \h.$ The
creation-annihilation operators therefore skip a level every time
they are applied. Finally notice that $b$ is proportional to
$\h^2,$ and therefore $\nu$ does not depend on $\h$. Hence the
energy levels are still grouped in pairs as $\h\rightarrow 0$.




The method of dressing chains (see \cite{VS, Ve}) can be used to
construct Hamiltonians for which we can calculate explicitly the
spectrum of energy. The idea is to consider a chain of
one-dimensional Hamiltonians $H_n$ that can be factored as

\beq \label{dressing}H_n=a_n a^\dag_n=a^\dag_{n+1}a _{n+1}+C_n\eeq

\noindent with the periodicity condition $H_{n+N}=H_n$ This method
is directly related to $2D$-superintegrability. Indeed, if one of
the Hamiltonians in the chain is $2D$-superintegrable, it is
possible to construct additional integrals for the whole family of
Hamiltonians, for if $[H_n(x,p_x)+H'(y,p_y),Q]=0,$ then
$$[H_{n+1}(x,p_x)+H'(y,p_y),a_{n+1} Q a_{n+1}^\dag]=0.$$

It turns out indeed that with $N=3,$ the first nontrivial case of
dressing chains, equations \eqref{dressing} were solved to give a
family of $2D$-superintegrable potential defined in terms of the
fourth Painlev\'e transcendant (see e.g.\cite{Ve}),
%
%
%
classified as (Q.17) in \cite{Gr}. Since Hamiltonians satisfying
dressing chains have interesting regularity and solvability
properties (see \cite{Ve}), the relation between the existence of
dressing chains and integrability deserves further study. Here we
will simply use one of the results of \cite{Ve} to give two
further examples of $2D$-superintegrable potentials with
interesting spectra.


Let us consider the potential
$$V=\frac{\h^2x^2}{8\a^4}+\frac{\h^2}{(x-\a)^2}+\frac{\h^2}{(x+\a)^2},$$

\noindent with $\a \in \RR^*,$ which is a special case of the
family of potentials classified as $(Q.17)$ in \cite{Gr}.

This potentials behaves at infinity like the harmonic oscillator,
but has two second-order poles at $x=\pm\a.$ The method of
dressing chains allows us in theory to calculate the spectrum for
this potential from that of the harmonic oscillator. Indeed, if we
write
\begin{equation}
\begin{split}
b=\left(p_x- \frac{i\h}{2\a^2}x-i \h f(x)\right)/\sqrt{2}\\
b^\dag=\left(p_x+ \frac{i\h}{2\a^2}x+i \h f(x)\right)/\sqrt{2}\\
\end{split}
\end{equation}

\noindent where $f(x)=-1/(x-\a)-1/(x+\a),$ we find
\begin{equation*}
\begin{split}
H_1=b^\dag b&=\frac{p_x^2}{2}+\frac{\h^2
x^2}{8\a^4}-\frac{5\h^2}{\a^2}\\
H_2=bb^\dag&=\frac{p_x^2}{2}+\frac{\h^2
x^2}{8\a^4}+\frac{\h^2}{(x-\a)^2}+\frac{\h^2}{(x+\a)^2}-\frac{3\h^2}{4\a^2}.
\end{split}
\end{equation*}

For every eigenvector $\ket{\phi}$ of $H_1,$ the function $b
\ket{\phi}$ is either zero, or an eigenvector of $H_2$.
Conversely, if we have an eigenfunction $\ket{\psi}$ of $H_2$ the
function $b^\dag \ket{\psi}$ is zero, or an eigenvector of $H_1.$
Since the spectrum and eigenvectors of $H_1$ are well-known, we
might get the impression that the problem for $H_2$ is solved.
This is not the case, since the eigenfunctions for $H_2$ obtained
by the application of $b$ on the analytic, bound states of the
harmonic oscillator $\phi_n(x)$ have a first-order pole at
$x=\pm\a,$ and are therefore not square-integrable, unless
$\phi_n(\pm\a)=0.$ Since this is the case only for $\phi_2,$ and
since $b\phi_2=0,$ we can not construct a single square integrable
bound state this way.


Let us therefore consider a hypothetic square-integrable ground
state $\psi$ of $H_2.$ We will now use an important result of
Veselov \cite{Ve} which states that all solutions of $H_2
\psi=E\psi$ (his result applies in fact to any Hamiltonian  of the
form $(Q.17)$) must be meromorphic. We can therefore develop
$\psi$ in Laurent series around $x=\a$ or $x=-\a.$ Since $\psi$ is
square integrable and meromorphic, it can not diverge at any
finite $x,$ hence the Laurent series are in fact Taylor series.

Let us now consider the Schr\"odinger equation $H_2 \psi=E\psi.$
Since $\psi$ is nonsingular, it must be proportional to
$(x-\a)^2(x+\a)^2.$ Therefore $b^\dag\psi,$ which is a
eigenfunction of $H_1,$ is proportional to $(x-\a)(x+\a),$ and
therefore does not diverge at $x=\pm\a.$ The function $b^\dag\psi$
is therefore an analytic eigenfunction of the harmonic oscillator.
If $b^\dag\psi=0,$ we can solve directly for $\psi$ and find

\begin{equation}
\psi(x)\propto \frac{\exp{\frac{x^2}{4 \a^2}}}{x^2-\a^2},
\end{equation}

\noindent which is not square integrable. If $b^\dag\psi\neq 0$
is square integrable, it is a standard solution of the harmonic
oscillator, and can be written as
$$h(x) e^{-\frac{x^2}{4 \a^2}}$$
\noindent where $h(x)$ is a polynomial in $x$.

We have just seen that in this case $b(b^\dag \psi)= E\psi$ cannot
be square-integrable, and therefore $E=0$ and $b^\dag\psi=\phi_2,$
the second excited level of the harmonic oscillator.




If we solve $b^\dag\psi=\phi_2,$
we find
$$\psi\propto \frac{x(x^2-\a^2)e^{-x^2/4 \a^2}-\sqrt{2\pi}\a^3e^{x^2/4
\a^2} \erf\left(\frac{x}{\sqrt{2}a}\right)}{x^2-\a^2}.$$

This function is not square integrable because of its behavior at
infinity and at $x=\pm\a.$

We have therefore proven that if $\psi$ is a square integrable
eigenstate of $H_2,$ the function $b^\dag\psi$ cannot be a square
integrable function.

Since it is an analytic eigenstate of the harmonic oscillator, we
know from undergraduate textbooks on quantum mechanics (i.e.
\cite{Ga}) that if $b^\dag\psi$ is not square integrable, it goes
to infinity as a polynomial times $e^{x^2/4\a^2}.$ If that is the
case, $bb^\dag\psi=E \psi$ cannot be square integrable at infinity
unless it is zero. Since we have already found the two states that
have $E=0$, we have shown that the Hamiltonian $H_2$ has no square
integrable eigenfunction. This Hamiltonian is therefore
equidistant, but in a quite degenerate manner. Interestingly
enough, if we consider the same Hamiltonian along the imaginary
axis, or equivalently with $i \a\in \RR,$ we find a real
Hamiltonian without singularities. We can this time find square
integrable bound states of $bb^\dagger$ from states of the
harmonic oscillator. We find indeed that the Hamiltonian
$bb^\dagger$ with $\a$ imaginary admits levels with $E=0$, and
with $E=(n+3)\h^2/(2\a)$ with $n\in\NN,$ but none at
$E=\h^2/(2\a)$ or $E=\h^2/\a.$ We therefore have an example where
the creation operator does not skip any level, but where the
spectrum is not exactly equidistant either.


The quantum $2D$-superintegrable potentials we studied here all
have in common with the harmonic oscillator that their spectrum
could be generated from a finite number of states by the
application of a creation operator. Their emission spectrum is
highly degenerate, which is exactly the kind of properties we
expect the quantum harmonic potentials to exhibit. Moreover, their
classical limit is the (isochronous) harmonic oscillator. Quantum
integrals yield classical integrals and it is therefore believed
that quantum superintegrable potentials have superintegrable
classical equivalents. Therefore quantum $2D$-superintegrable
Hamiltonians are expected to have $2D$-superintegrable classical
equivalents. This provides an additional indication that the use
of $2D$-superintegrability to generalize isochronicity or
harmonicity is appropriate.

%

All this also provides a new insight on the general significance
of superintegrability in quantum mechanics. Superintegrability has
already been related to superseparability and exact solvability in
quantum mechanics \cite{TTW}. We have noticed in \cite{Gr} that
quantum $2D$-superintegrable potentials known today are all
solutions to equations having the Painlev\'e property. Moreover,
the result by Veselov \cite{Ve}, and the relation we established
between superintegrability and dressing chains tells us that the
eigenstates for these Hamiltonians are often meromorphic in the
complex plane.  Our paper shows that superintegrability is also
related to more physical properties, both in classical and quantum
mechanics.

We have shown rigorously the equivalence between
$2D$-superintegrable potentials with a minimum and isochronous
ones. We also showed that this characterization of isochronous
potentials could easily be transposed into more general contexts,
such as potentials without a local minimum, or, more importantly,
to quantum mechanics. $2D$-integrability can be easily generalized
to Hamiltonians separable in different set of coordinates, or else
in higher dimensions. Considering maximally superintegrable
three-dimensional potentials of the form $V_1(x,y)+V_2(z),$ for
example, allows us to deal with $2D$ isochronous potentials. The
results we obtained together with this variety of possible
generalizations demonstrates the fruitfulness of the approach
presented here.

The author thanks P. Winternitz and C. Rousseau for useful
discussions. This work was supported by NSERC of Canada.


\end{document}